# Sub-femtosecond absolute timing precision with a 10 GHz hybrid photonic-microwave oscillator

**T.M. Fortier, C.W. Nelson, A. Hati, F. Quinlan, J. Taylor, H. Jiang, C.W. Chou, T. Rosenband, N. Lemke, A. Ludlow, D. Howe, C.W. Oates and S.A. Diddams**[*]

*National Institute of Standards and Technology, Time and Frequency Division*

*325 Broadway, MS 847 Boulder, CO 80305*

We present an optical-electronic approach to generating microwave signals with high spectral purity. By circumventing shot noise and operating near fundamental thermal limits, we demonstrate 10 GHz signals with an absolute timing jitter for a single hybrid oscillator of 420 attoseconds (1Hz – 5 GHz).

Stable periodic signals are central to a wide class of physical measurements in basic and applied science, where precise knowledge of the phase of an oscillator waveform can translate to precision determination of distance and time intervals. As measurement requirements become more stringent, there is a corresponding need for oscillators that better approximate ideal periodic signals. This is particularly true for microwave signals on a timescale less than 1 second where lower oscillator noise enables higher sensitivity measurement in advanced radar[1], high-speed sampling[2], timing synchronization in large-scale scientific experiments[3,4], and precision atomic spectroscopy[5]. Recently, photonic techniques have provided a new approach to generating stable microwave signals via optical frequency division[6-9]. As with any photonic approach, the spectral purity of the generated microwaves is fundamentally limited at shorter times by photodetection shot noise[10-13]. In this work, we circumvent this limit by disciplining a room temperature sapphire loaded cavity oscillator (SLCO)[14] to the microwave signal generated via optical frequency division. With this technique we generate a 10 GHz microwave signal with a 500-fold (20-fold) improvement in timing precision as compared to one of the best conventional microwave (laboratory-based photonic) oscillators.

The experimental setup used for generation and characterization of two independent hybrid microwave sources is depicted in Figure 1. Our optical approach to microwave generation relies on division of the frequency $\nu_{opt}$ of a CW laser stabilized to a high quality factor ($Q \sim 10^{11}$) optical reference cavity[15,16]. Division from ~500 THz to the microwave domain is accomplished with an optical frequency comb divider (OFCD) that is based on a 1 GHz Kerr-lens mode-locked Ti:sapphire laser.[17] (We note the hybrid oscillator technique would work equally well using an OFCD based on an Er:fiber laser that has demonstrated comparable phase noise close to carrier[18]). This pulsed laser has an optical spectrum that is composed of approximately $3 \times 10^5$ optical frequencies, which are exactly spaced by the repetition rate $f_r = v_g/L$, where $L$ is the round trip path length in the mode-locked laser cavity and $v_g$ is the group velocity of the light in the cavity. Control of $f_r$, in addition to the carrier-envelope offset frequency of the pulses, allows us to phase lock a single optical mode to that of the cavity stabilized reference laser[17,18]. This transfers the frequency stability of the CW laser to a timing stability in the pulse laser repetition rate, $f_r$. Significantly, in frequency division, the power spectral density of phase noise on $\nu_{opt}$ is reduced by a factor of $N^2$, where $N=50,000$ is the ratio of optical and microwave frequencies. Harmonics of $f_r$, up to the cut-off bandwidth of the detector, are generated via photodetection of ~12 mW of the pulsed laser light with a power of -8 dBm obtained in the 10$^{th}$ harmonic. After filtering and amplification, a low-noise 10 GHz signal is available for characterization and use in other applications.

---

[*] email: *sdiddams@boulder.nist.gov*

Although this optical approach to microwave generation results in exceptional phase stability close to carrier, for offset frequencies >10 kHz we encounter a limit in the phase noise near -153 dBc/Hz that results from fundamental photodetector shot noise[9,10]. Shot noise typically dominates over thermal noise for >1mA of detected photocurrent. Under these circumstances one would expect the signal-to-noise ratio (SNR) to be proportional to the average photocurrent. The effect of shot noise on the SNR is exacerbated by the saturation of the peak photocurrent and associated microwave power at higher harmonics of $f_r$. Saturation arises from nonlinear effects in the photodiode[18] and leads to a decrease in the detector bandwidth, such that the SNR of the higher harmonics of $f_r$ no longer improves with increasing photocurrent[11]. This problem can be alleviated to some extent with improved photodiode design[20,21] and by repetition rate multiplication[11-13] of the input source, thereby reducing the peak optical power and better concentrating the microwave power in the desired harmonic (e.g. 10 GHz).

The 10 GHz signal from a high power commercial SLCO, on the other hand, can reach significantly lower thermal noise limited floors at offset frequencies >1 MHz[14]. In the measurements presented here we demonstrate a combined noise floor for our SLCOs of -190 dBc/Hz at frequency offsets > 1 MHz. To generate the necessary microwave power to achieve a comparable noise floor via our optical division approach, one would require a single photodiode capable of handling several amperes of photocurrent with ≥10 GHz bandwidth. While the SLCO operates with high circulating power, it has a significantly lower quality factor ($Q$) than the best passive optical cavity. As a result it produces a 10 GHz waveform with low noise at high Fourier frequencies, but one that is less stable than that generated via optical means at offset frequencies < 5 kHz.

To combine the best phase noise characteristics of both optical and microwave oscillators, the 10 GHz microwave signal (+15 dBm) from the SLCO is phase locked to that from the OFCD using an electronic phase lock with a 5 kHz loop bandwidth, as shown in Fig. 1. Determination of the absolute phase noise of this hybrid 10 GHz oscillator is accomplished via comparison with a second nearly identical hybrid system. Correlated noise in the hybrid systems at frequencies of 1 Hz - 10 kHz is expected to be negligible, as the optical reference cavities for the two oscillators are located in laboratories separated by hundreds of meters. Frequency stable light from the optical cavities is transported by noise-cancelled fiber[9] to the two OFCDs, which are located in a common lab, but on separate isolated platforms. The optical pulse trains from the OFCDs are connected to the photodiodes, SLCO's and the measurement system via additional short optical fibers. As seen in Figure 1, the outputs of the two hybrid systems are compared using a cross-correlation spectral measurement[22]. This technique allows for suppression of the uncorrelated electronic noise of amplifiers and mixers in the measurement system by $\sqrt{m}$, where $m$ is the number of measurements that are averaged. The relative frequency drift between the optical reference cavities is sufficiently low such that the appropriate quadrature phase relationship of the 10 GHz reference signals is maintained in the measurement system over extended periods without the need for a slow feedback servo.

Figure 2 summarizes our measurements of the single-sideband phase noise, $L(f)$, of the 10 GHz microwave signals from the photonic and sapphire oscillators as well as the output of the hybrid oscillators. All curves represent the full noise contributed by a pair of oscillators. As seen in Fig. 2c, the noise spectrum of the hybrid signal is the combination of the high frequency spectrum from the SLCO microwave source (Fig. 2a) and the low frequency noise spectrum from the photonic oscillator (Fig. 2b). Aside from allowing us to circumvent the photonic generator shot noise floor, the hybrid oscillator also eliminates the excess phase noise that results from amplitude-to-phase conversion of the OFCD relative intensity noise in the photodetector[23,24]. This effect is most noticeable for frequencies 50-500 kHz and can be seen in Fig 2 (b) as a broad band bump in the phase noise spectrum that is centered at 200 kHz. At lower offset frequencies the excess phase noise from amplitude-to-phase conversion is significantly lower than the phase noise of the photonic oscillator itself.

Integration of the phase noise of Fig 2c results in a combined absolute timing jitter for the two hybrid 10 GHz oscillators of 305 attoseconds (1Hz – 1 MHz). Extending the integration to the Nyquist frequency of 5 GHz at the SLCOs thermal noise floor of -190 dBc/Hz increases this timing jitter to only 590 attoseconds. Assuming that both hybrid oscillators contribute equally, the single oscillator timing jitter would be 420 attoseconds. For comparison, the absolute timing jitter for the pairs of SLCOs and OFCDs are 300 fs and 11.4 fs (1 Hz – 5 GHz), respectively. In the case of the OFCDs, we have assumed the shot noise floor extends to 10 MHz, at which point a bandpass filter could be employed to reduce the noise floor to the thermal limit of approximately -163 dBc/Hz. Thus, locking the SLCO to the OFCD reduces the timing jitter of the free-running SLCO by a factor of 500.

Amplitude noise is also an important parameter of any precision oscillator. Presented in Figure 3 is the amplitude noise on the 10 GHz tone for signals generated via the hybrid generator. For this measurement, we again use a cross-correlation measurement, where the output of a single oscillator is split and detected with two microwave diodes whose outputs are sent to two baseband amplifiers. The outputs of the amplifiers are then cross-correlated with an FFT spectrum analyzer. Integration of the noise spectrum of the hybrid signal (Fig. 3) over the bandwidth of 10 Hz – 5 GHz results in an integrated intensity noise of 0.009%. In this case, we have assumed a noise floor of -178 dBc/Hz at frequencies above 10MHz. The integrated noise of the hybrid oscillator is the same as that of a free running SLCO and is a factor of 7 lower than the AM noise of an OFCD photonic generator.

In summary, our combination of the best of photonic and electronic microwave technologies has produced an electromagnetic signal with exceptional spectral purity in the important X-band microwave region. In doing so we take advantage of the benefits of both technologies while circumventing fundamental photodetection shot noise to achieve a thermally-limited noise floor. We anticipate this hybrid approach will favorably impact scientific and technical applications that require time interval and distance measurement with the highest precision over timescales extending from hundreds of picoseconds to seconds. Extension of sub-femtosecond timing precision to significantly longer timescales can be achieved by long-term steering of the hybrid oscillator output to an atomic reference[25,26].

**REFERENCES**

[1] J. A. Scheer and J. Kurtz, *Coherent radar performance estimation*, (Artech House, 1993).

[2] R. H. Walden, IEEE J. Sel. Comm. **17**, 539 (1999).

[3] J. Kim, J. A. Cox, J. Chen and F. X. Kartner, *Nat. Photon.* **2**, 733 (2008).

[4] S. Doeleman in *Frequency standards and metrology: Proceedings of the 7th symposium*, edited by L. Maleki (World Scientific, 2009), pp 175-183.

[5] G. Santarelli, Ph. Laurent, P. Lemonde, A. Clairon, A. G. Mann, S. Chang, A. N. Luiten, and C. Salomon, Phys. Rev. Lett. **82**, 4619 (1999).

[6] S. A. Diddams, A. Bartels, T. M. Ramond, C. W. Oates, S. Bize, E. A. Curtis, J. C. Bergquist, and L. Hollberg, IEEE Journ. Select. Topics Quant. Electron. **9**, 1072 (2003).

[7] A. .Bartels, S.A. Diddams, C.W. Oates, G. Wilpers, J. C. Bergquist, W. Oskay, L. Hollberg, Opt Lett. **30**, 667 (2005).

[8] J. Millo, M. Abgrall, M. Lours, E. M. L. English, H. Jiang, J. Guéna, A. Clairon, M. E. Tobar, S. Bize, Y. Le Coq, and G. Santarelli, Appl. Phys. Lett. **94**, 141105 (2009).

[9] T.M. Fortier, M.S. Kirchner, F. Quinlan, J. Taylor, J.C. Bergquist, T. Rosenband, N. Lemke, A. Ludlow, Y. Jiang, C.W. Oates, and S.A. Diddams. *Nat. Photon.* **5**, 425 (2011).

[10] J.J. McFerran, E.N. Ivanov, G. Wilpers, C.W. Oates, S.A. Diddams, and L. Hollberg, Electron. Lett. **41**, 36 (2005).

[11] S. A. Diddams, M. Kirchner, T. Fortier, D. Braje, A. M. Weiner, and L. Hollberg, Opt. Exp. **17**, 3331 (2009).

[12] A. Haboucha, W. Zhang, T. Li, M. Lours, A. N. Luiten, Y. Le Coq, and G. Santarelli, Opt. Lett **36**, 3654 (2011).

[13] H. Jiang, J. Taylor, F. Quinlan, T. Fortier and S. A. Diddams, IEEE Photonics Journal **3**, 1004 (2011).

[14] E. N. Ivanov and M. E. Tobar, IEEE Trans. Ultrason. Ferro and Freq. Control **56**, 263 (2009).

[15] B.C. Young, F.C. Cruz, J.C. Bergquist, and W.M. Itano, *Phys. Rev. Lett.* **82**, 3799 (1999).

[16] Y. Jiang, A. Ludlow, N. Lemke, J. Sherman, J. von Stecher, R. Fox, L.-S. Ma, A.M. Rey, and C. Oates, *Nat. Photon.* **5**, 2 (2011).

[17] T.M. Fortier, A. Bartels, and S.A. Diddams, Opt. Lett. **31**, 1011 (2006).

[18] F. Quinlan, T.M. Fortier, M.S. Kirchner, J.A. Taylor, M.J. Thorpe, N. Lemke, A.D. Ludlow, Y. Jiang, and S.A. Diddams, Opt. Lett. **36**, 3260 (2011)

[19] K J. Williams, R. D. Esman, and Mario Dagenais, IEEE J. Lightwave Tech. **14**, 84 (1996).

[20] A. Joshi, S. Datta, and D.Becker, IEEE Photon. Technol. Lett., **20**, 1500 (2008).

[21] Z. Li, H. Pan, H. Chen, A. Beling, and J. C. Campbell, IEEE J. Quant. Electron. **46**, 626 (2010).

[22] Warren F. Walls, *Proc. IEEE Freq. Contr. Symp.*, 1992, pp. 257-261.

[23] J. A. Taylor, S. Datta, A. Hati, C. Nelson, F. Quinlan, A. Joshi, and S. A. Diddams, IEEE Photon. J. **3**, 140 (2011).

[24] W. Zhang, T. Li, M. Lours, S. Seidelin, G. Santarelli, and Y. Le Coq, Appl. Phys. B **106**, 301 (2012).

[25] T. Rosenband, P.O. Schmidt, D. Hume, W.M. Itano, T. Fortier, J. Stalnaker, K. Kim, S.A. Diddams, J. Koelemeij, J.C. Bergquist, and D.J. Wineland, Phys. Rev. Lett. **98**, 220801-4 (2007)

[26] N. Lemke, A.D. Ludlow, Z. Barber, T. Fortier, S.A. Diddams, Y. Jiang, S.R. Jefferts, T.P. Heavner, T.E. Parker, and C.W. Oates, Phy. Rev. Lett, **103**, 063001-4 (2009).

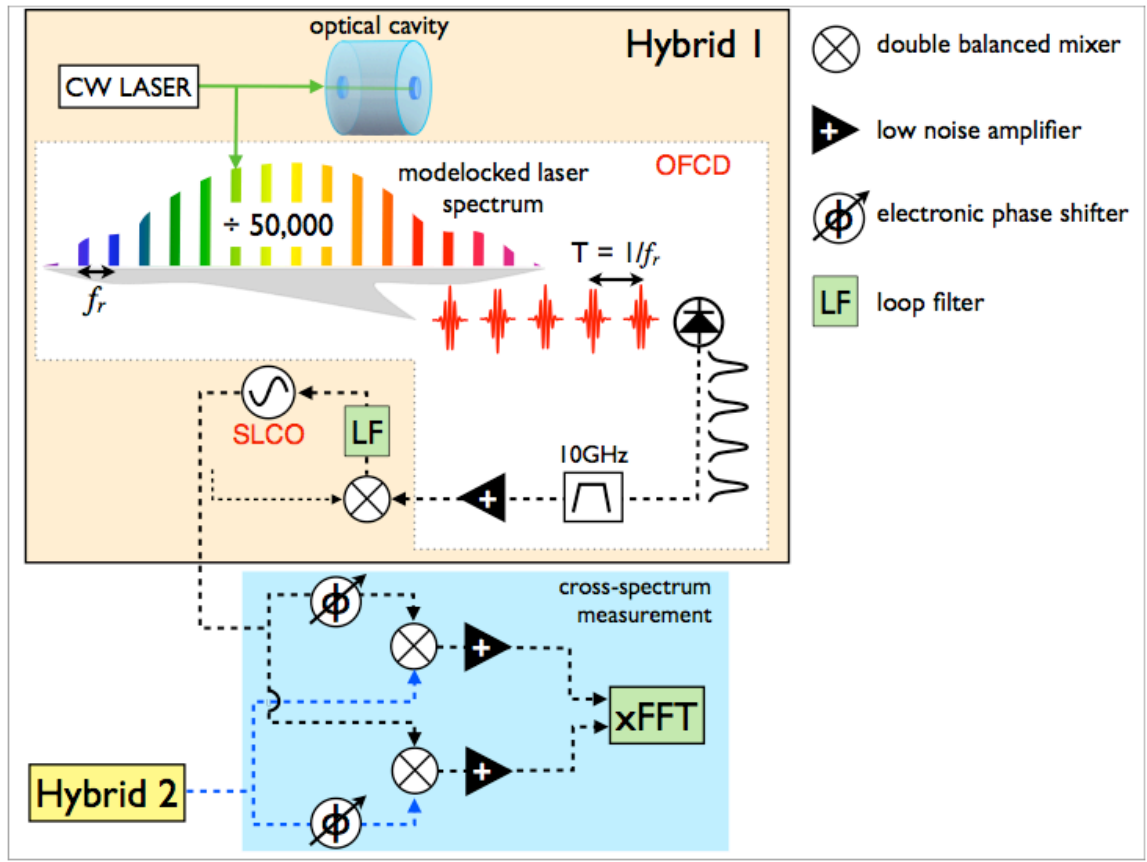

FIG 1. Generation and characterization of 10 GHz signals from hybrid oscillators. Each hybrid oscillator is formed from the combination of laser-based and sapphire dielectric microwave oscillators. The optical frequency of the stable laser is divided down to the microwave domain with an optical frequency comb divider (OFCD). Photodetection of the laser pulse train results in a train of electronic pulses. The 10$^{th}$ harmonic of the laser repetition rate at 10 GHz is filtered with a bandpass filter and amplified. The 10 GHz microwave signal from an SLCO is phase locked to this electronic signal with a loop bandwidth of 5kHz. Phase noise characterization of two independent, but similar, hybrid oscillators is accomplished via a cross spectrum measurement.

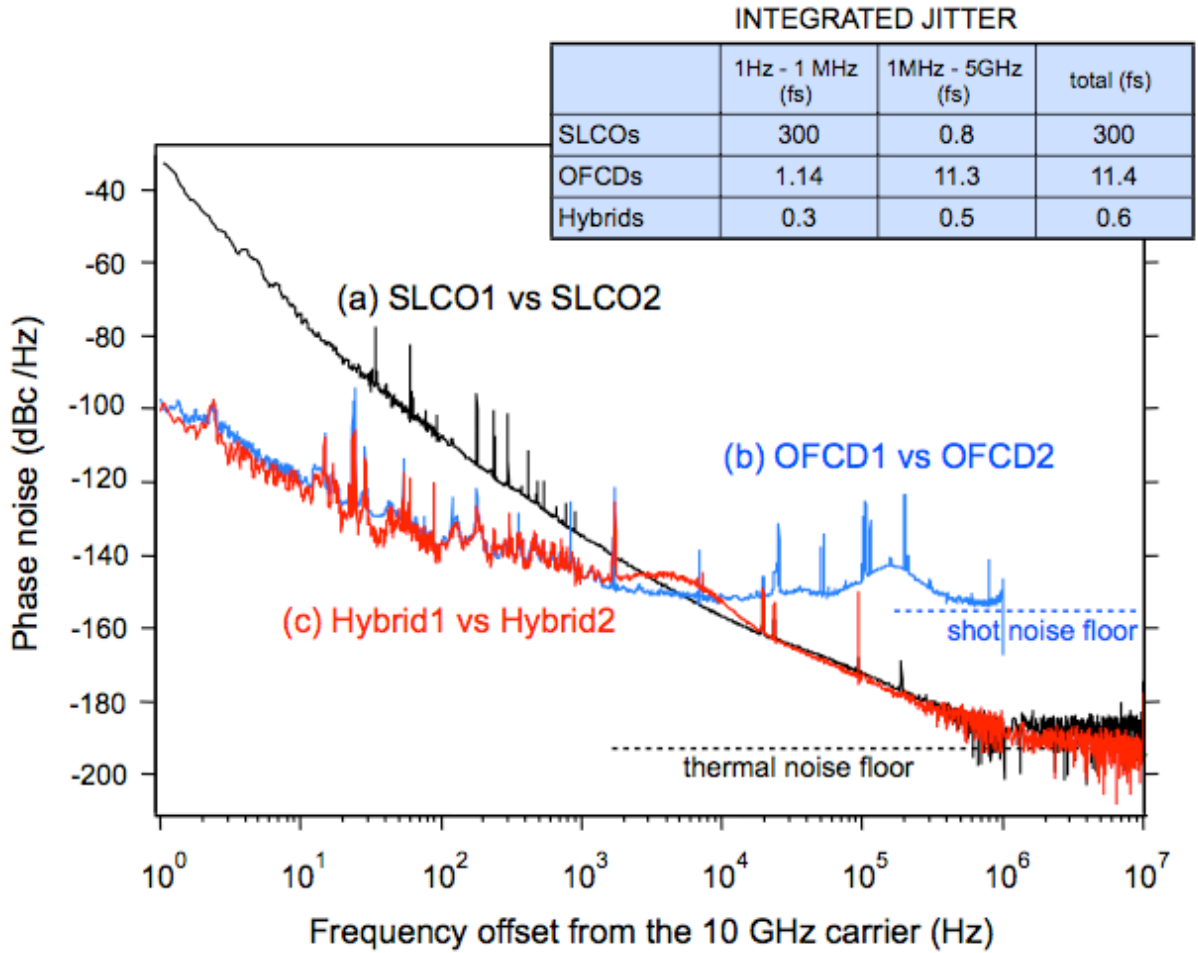

FIG 2. Power spectral density of the phase noise on the 10 GHz signal from (a) microwave sapphire loaded cavity oscillator (SLCO), (b) optical oscillator with a frequency-comb-based optical frequency divider (OFCD), and (c) the hybrid oscillator based on the combination of the microwave and optical devices. A phase noise power of -190 dBc/Hz, which is near the thermal noise floor level of the SLCO, indicates that in a 1 Hz bandwidth the noise power is 19 orders of magnitude below the 10 GHz carrier. To measure such a noise floor required 100,000 averages acquired in fifteen minutes. The inset table provides the integrated timing jitter of the oscillators. All curves and timing jitter values represent the measured noise of a pair of near-identical oscillators.

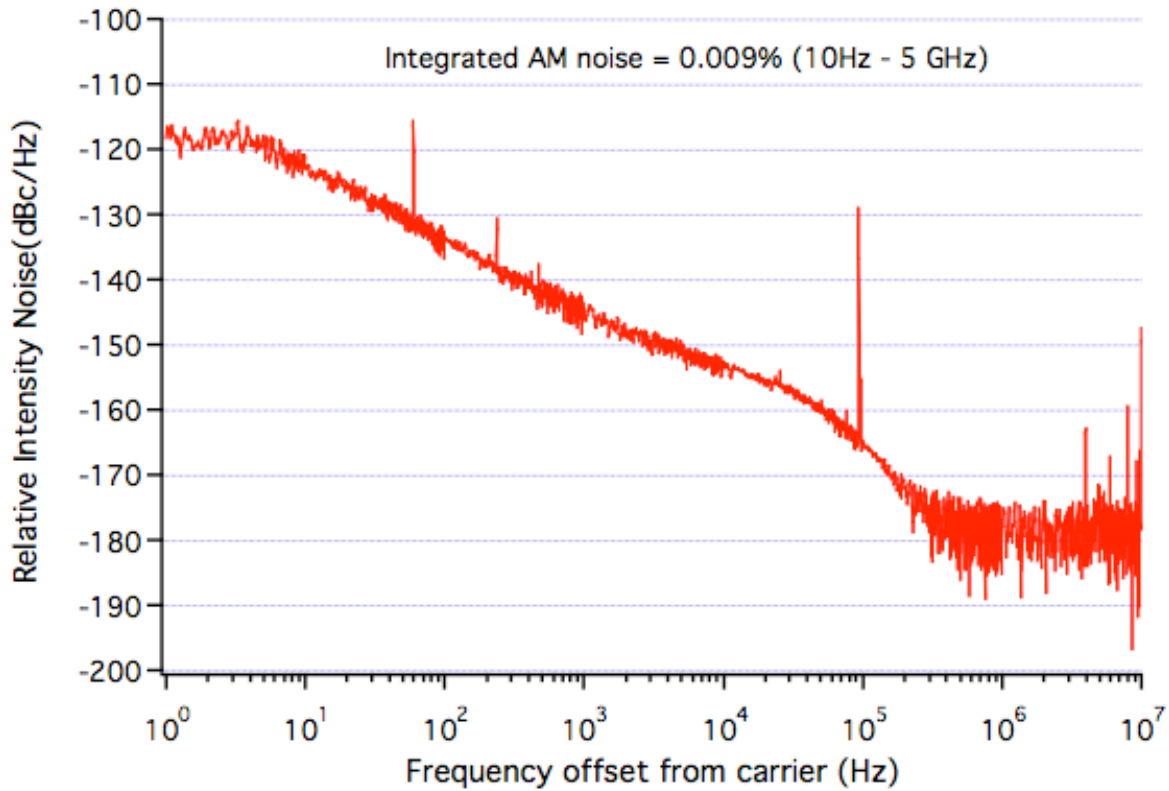

FIG 3. AM noise measurement of the 10 GHz signal from a hybrid oscillator. The observed flattening near 1 Hz is an artifact resulting from AC coupling of the baseband amplifier. The AM noise floor for the hybrid oscillator was limited by the number of averages in the cross spectral measurement.